\documentclass[11pt]{article}
\usepackage{graphicx, caption, subfigure}
\makeatletter
\newcommand{\singlespacing}{\let\CS=\@currsize\renewcommand{\baselinestretch}{1.0}\tiny\CS}
\newcommand{\doublespacing}{\let\CS=\@currsize\renewcommand{\baselinestretch}{1.5}\tiny\CS}

\begin{document}
\title{Production of J/$\Psi$-Particles at RHIC and LHC energies: An Alternative `Psi'-chology}
\author{P.
Guptaroy$^1$\thanks{e-mail: gpradeepta@rediffmail.com}, Goutam
Sau$^2$\thanks{e-mail: sau$\_$ goutam@yahoo.com}, S. K. Biswas$^3$\thanks{e-mail:
sunil$\_$biswas2004@yahoo.com}   \&
S. Bhattacharyya$^4$\thanks{e-mail: bsubrata@isical.ac.in
}\\
{\small $^1$ Department of Physics, Raghunathpur College,}\\
 {\small P.O.: Raghunathpur 723133,  Dist.: Purulia(WB), India.}\\
{\small $^2$ Beramara Ram Chandrapur High School,}\\
 {\small South 24-Parganas, 743609(WB), India.}\\
 {\small $^3$ West Kodalia Adarsha Siksha Sadan, New Barrackpore,}\\
 {\small Kolkata-700131, India.}\\
 {\small $^4$ Physics
and Applied Mathematics Unit(PAMU),}\\
 {\small Indian Statistical Institute,}\\
 {\small 203 B. T. Road, Kolkata - 700108, India.}}
\date{}
\maketitle
\bigskip
\bigskip
\begin{abstract}
We attempt here to understand successfully some crucial aspects of $J/\Psi$-production in some high energy nuclear
collisions in the light of a non-standard framework outlined in
the text. It is found that the
results arrived at with this main working approach here is fairly in
good agreement with both the measured data and the results obtained
on the basis of some other models of the `standard' variety. Impact and implications
of this comparative study have also been precisely highlighted in
the end.
\end{abstract}

 \newpage
 \doublespacing
 \section{Introduction}
 The study of the $J/\Psi$-mesons in ultra-relativistic heavy ion
collisions has consistently been considered to be a potentially powerful tool for studying the properties of the hypothetical
`hot and dense matter' created in these collisions. Predictions about the $\textit{suppression}$ of $J/\Psi$-meson production in the nuclear collision, at
certain stages \cite{matsui} and an anomalous $\textit{enhancement}$ at certain
other stage as well \cite{thews} are always treated as very powerful diagnostics. The deciding factor, in terms of the Standard Model (SM), in both the cases, is the number of charm-anticharm ($c{\bar c}$) pairs ($N_{c{\bar c}}$) created in the early stage of hard
parton collisions in $A + A$ (or in $A + B$) reactions. On the basis of it, one might arrive at the assumptions of the $J/\Psi$-$\textit{suppression}$ (at $ N_{c{\bar c}} < 1$) and the $J/\Psi$-$\textit{enhancement}$ (at $N_{c{\bar c}} > 1$) effects.
\par
 In the past \cite{pgr05}-\cite{pgr09} we had constantly refrained from giving
such undue importance to any controversy about either suppression/enhanced suppression or an enhancement of $J/\Psi$ mesons in reality. Rather our primary
intent would centre around understanding and interpreting the nature
of data of inclusive cross-sections and some other important
observables of $J/\Psi$ mesons in BNL-RHIC and CERN-LHC experiments. In
fact, we have an altogether different and heretic view (which will be outlined very briefly in the next section) about the
mechanism of $J/\Psi$ production in high energy particle and nuclear
collisions.
\par
We define our objectives here as: 1) to explain the main and major
features of the latest data on $J/\Psi$-production in BNL-RHIC and CERN-LHC
experiments from the proposed alternative approach built up by us in
a set of previous works done in the both the remote and recent past
\cite{bhat881}-\cite{pgr10} and 2) to compare our model-based calculations with some other competing models.
\par
Our plan of work presented here is as follows. In section 2, we provide brief outlines of
the models chosen for this study. The section 3 provides very brief outlines of the models which are founded on the gantlets of assumptions of the `Standard'-model and which have been reckoned with for the sake of comparison.
In section 4 we give the results and general discussion. And in the
last section (section 5) we offer the final remarks-cum-conclusions.
 \section{The Main Approach: An Outline}
 In the present work, we will make use of a theoretical model to interpret some of the latest and topical observables of the J/$\Psi$ production which were measured and reported
by the different groups in the Relativistic Heavy Ion Collider (RHIC) and Large Hadron Collider (LHC) experiments in the recent past.
The model has a modest degree of dynamical basis and some prior check-ups with
data \cite{pgr05}-\cite{pgr09}. It is called here as the sequential chain model (SCM).
\par
The outline and features of the model we use here are obtained, in
the main, from our previous works \cite{pgr05}-\cite{pgr09}. According to this model, high energy hadronic interactions
boil down, essentially, to the pion-pion interactions; as the
protons are conceived in this model as
$p$=($\pi^+$$\pi^0$$\vartheta$), where $\vartheta$ is a spectator
particle needed for the dynamical generation of quantum numbers of
the nucleons. The multiple production of $J/\Psi$-mesons in a high
energy proton-proton collisions is described in the following way.
The secondary $\pi$-meson or the exchanged $\varrho$-meson emit a
free $\omega$-meson and pi-meson; the pions so produced at high
energies could liberate another pair of free $\varrho$ and trapped
$\omega$-mesons (in the multiple production chain). These so-called
free $\varrho$ and $\omega$-mesons decay quite a fast into photons
and these photons decay into $\Psi$ or $\Psi'$ particles, which,
according to this alternative approach is a bound state of $\Omega
\bar{\Omega}$ or $\Omega' \bar{\Omega}'$ particles.
\par
The inclusive cross-section of the $\Psi$-meson produced in the $pp$ collisions given by
\begin{equation}\displaystyle{
E\frac{d^3\sigma}{{dp}^3}|_{p+p\rightarrow{{J/\Psi}+X}}  \cong
C_{J/\Psi}\frac{1}{p_T^{N_R}}\exp(\frac{-5.35(p_T^2+m^2_{J/\Psi})}{<n_{J/\Psi}>^2_{pp}(1-x)})
\exp(-1.923{<n_{J/\Psi}>_{pp}}x),}
\end{equation}
where the expression for for average multiplicity for $\Psi$-particles in
$pp$ scattering would be given by
\begin{equation}\displaystyle{
<n_{J/\Psi}>_{pp} ~~~ = ~~~ 4\times10^{-6}s^{1/4}.}
\end{equation}
In the above expression, the term $|C_{J/\Psi}|$ is a
normalisation parameter and is assumed here to have a value $\cong
0.09$ for Intersecting Storage Ring(ISR) energy, and it is
different for different energy and for various collisions. The
terms $p_T$, $x$ and $m_{J/\Psi}$ represent the transverse
momentum, Feynman Scaling variable and the rest mass of the
$J/\Psi$ particle respectively. Moreover, by definition, $x ~ = ~
2p_L/{\sqrt s}$ where $p_L$ is the longitudinal momentum of the
particle. The $s$ in equation (2) is the square of the c.m. energy.
\par
The second term in the right hand side of the equation (1), the
constituent rearrangement term arises out of the partonic
rearrangements inside the proton. It is established that
hadrons (baryons and mesons) are composed of few partons. These rearrangements mean
undesirable loss of energy , in so far as the production mechanism
is concerned. The choice of ${N_R}$ would depend on the following
factors: (i) the specificities of the interacting projectile and
target, (ii) the particularities of the secondaries emitted from a
specific hadronic or nuclear interaction and (iii) the magnitudes of
the momentum transfers and of a phase factor (with a maximum value
of unity) in the rearrangement process in any collision. The parametrisation is to be
done for two physical points, viz., the amount of momentum transfer
and the contributions from a phase factor arising out of the
rearrangement of the constituent partons. Collecting and combining
all these, we propose the relation to be given by \cite{pgr072}
\begin{equation}\displaystyle
N_R=4<N_{part}>^{1/3}\theta,
\end{equation}
where $<N_{part}>$ denotes the average number of participating
nucleons and $\theta$ values are to be obtained phenomenologically
from the fits to the data-points \cite{bhat882}.
\par
In order to study a nuclear interaction of the type
$A+B\rightarrow Q+ x$, where $A$ and $B$ are projectile and target
nucleus respectively, and $Q$ is the detected particle which, in the
present case, would be $J/\Psi$-mesons, the SCM
has been adapted, on the basis of the suggested Wong \cite{wong} work to the
Glauber techniques. The inclusive
cross-sections for $J/\Psi$ production in different nuclear
interactions of the types $A+B\rightarrow J/\Psi+ X$ in the
light of this modified Sequential Chain Model (SCM) can then be written
in the following generalised form as:
\begin{equation}\displaystyle
{E{\frac{d^3\sigma}{dp^3}}|_{A+B\rightarrow{J/\Psi}+ X}=
a_{J/\Psi} {p_T}^{-N_R} \exp(-c(p_T^2+m^2_{J/\Psi}))
\exp(-1.923{<n_{J/\Psi}>_{pp}x)}}.
\end{equation}
 where $a_{J/\Psi}$, $N_R$ and $c$ are
the factors to be calculated under certain physical constraints. The
set of relations to be used for evaluating the parameters
$a_{J/\Psi}$  is given below.
\begin{equation}\displaystyle{
a_{J/\Psi}=C_{J/\Psi}{\frac{3}{2\pi}}{\frac{(A \sigma_B + B
\sigma_A)}{\sigma_{AB}}}
{\frac{1}{1+a'(A^{1/3}+B^{1/3})}}}
\end{equation}
Here, in the above set of equations, the third factor gives a
measure of the number of wounded nucleons i.e. of the probable
number of participants, wherein $A\sigma_B$ gives the probability
cross-section of collision with `$B$' nucleus (target), had all
the nucleons of $A$ suffered collisions with $B$-target. And
$B\sigma_A$ has just the same physical meaning, with $A$ and $B$
replaced.
\par
Besides, in expression (5), the fourth term is a physical factor
related with energy degradation of the secondaries due to multiple
collision effects. The parameter $a'$ occurring in eqn.(5) above
is a measure of the fraction of the nucleons that suffer energy
loss. The maximum value of $a'$ is unity, while all the nucleons
suffer energy loss. This $a'$ parameter is usually to be chosen
\cite{wong}, depending on the centrality of the collisions and the
nature of the secondaries.
\par
The ``$a$" factor in the expression (5) accommodates a wide range
of variation because of the existence of the large differences in
the in the normalizations of the $J/\Psi$ cross-sections for
different types of interactions.
\section{Some Competing Models: The Brief Outlines}
\subsection{The Gluon Saturation Approach}
In this approach \cite{adare11}, it is assumed
that the nuclear wave functions in very high-energy nuclear
collisions can be described by the Color Glass Condensate
(CGC). The primary effect is the suppression
of $J/\Psi$ production and narrowing of the rapidity distribution
due to saturation of the gluon fields in heavy
ion collisions relative to $p + p$ collisions. In addition,
the production mechanism is modified from $p + p$ such
that the multigluon exchange diagrams are enhanced in heavy ion reactions. It
should be noted that this model does not include any hot
medium effects, but does have a free parameter for the
overall normalization factor.
\subsection{The Quark Coalescence Model (QCM)}
The Quark Coalescence Model (QCM) \cite{kahana}, is
a two-stage simulation. In stage I the incoming target and
projectile nucleon interactions are tracked, a fluid of pre-hadrons
are formed and in stage II the produced pre-hadrons interact and
decay in a standard relativistic cascade model. The time history of
all the collisions recorded in stage I sets up the geometry and
initial conditions for stage II. Basic inputs for the simulation are
measured hadron-hadron crosssections, rapidity, transverse momentum
and particle multiplicity distributions, of which the last are
conforming to KNO scaling. The basic assumption of the model is
coalescence of charm quarks and anti-quarks into charmonia ($c{\bar
c} \rightarrow$ charmonium), which can also be viewed as heavy
pre-hadrons in stage I. Only a small percentage of charm quarks are
expected to coalesce into bound charmonia i.e. $J/\Psi$, the
remainder of such quarks appear ultimately as open charm mesons. At
RHIC energies comover suppression of directly produced charmonia is
expected to be large, due to the increased particle numbers and
densities.
\subsection{Double Color Filter-Oriented Approach}
 A mechanism called double color filtering for
$c{\bar c}$ dipoles \cite{kope}, makes nuclei significantly more
transparent in $AA$ than $pA$ collisions. The assumption of this
model is as follows: In the c.m. of nuclear collision the nuclear
disks passing through each other leave behind a cloud of radiated
gluons creating a dense matter, which the $J/\Psi$ propagates
through. In this reference frame the $J/\Psi$ full momentum is
$p_T$, which ranges from zero to several GeV in RHIC data. Such a
low energy $c{\bar c}$ dipole develops the $J/\Psi$ wave function
pretty fast, during time $t_f < 0.5$ fm, which is about the time
scale of the medium creation. Thus, what is propagating through the
medium is not a small $c{\bar c}$ dipole, but a fully formed
$J/\Psi$. The observed nuclear effects in $J/\Psi$  production in
$AA$ collisions is interpreted as a combination of final state
interaction (FSI) of $J/\Psi$ in the dense medium, and the initial
state interaction (ISI) effects in production of $J/\Psi$ caused by
multiple interactions of the colliding nuclei.
\section{The Results}
Now let us proceed to apply the chosen model to interpret some recent experimental results of $J/\Psi$-production reported by various groups for different collisions like $p+p$, $d+Au$, $Cu+Cu$,  $Au+Au$ at RHIC and $p+p$, $Pb+Pb$ at LHC. Here, the main observables are the invariant yields,
rapidity distributions and the nuclear modification factors which would come under purview of the present work.
\subsection{$J/\Psi$ (total) Crosssections and Rapidity Distribution in $p+p$ Interactions}
As the psi-productions are generically treated rightly as the
resonance particles, the standard practice is to express the
measured $J/\Psi$ (total) crosssections times branching ratio to
muon or electrons, i.e.for lepton pairs , that is by
$B_{ll'}\sigma^{J/\Psi}_{p+p}$.
\par
By using expression (4) we arrive at the expressions for the
differential cross-sections for the production of $J/\Psi$-mesons in
the mid and forward-rapidities (i.e. $|y|<0.35$ and $1.2<|y|<2.2$
respectively) in $p+p$ collisions at $\sqrt{s_{NN}}$=200 GeV at
RHIC.
\begin{equation}\displaystyle{
\frac{1}{2\pi p_T} B_{ll'} \frac{d^2\sigma}{dp_Tdy}|_{p+p\rightarrow
J/\Psi+X} = 6.1 p_T^{-1.183} \exp[-0.13(p_T^2+9.61)]
~~~~ for ~~|y|<0.35,}
\end{equation}
and
\begin{equation}\displaystyle{
\frac{1}{2\pi p_T} B_{ll'} \frac{d^2\sigma}{dp_Tdy}|_{p+p\rightarrow
J/\Psi+X} = 6.5 p_T^{-1.183} \exp[-0.16(p_T^2+9.61)]
~~~~ for ~~1.2<|y|<2.2.}
\end{equation}
For deriving the expressions (6) and (7) we have used the relation
$x\simeq \frac{2p_{Zcm}}{\sqrt s}= \frac{2m_T \sinh y_{cm}}{\sqrt
s}$ \cite{pdg}, where $m_T$ , $y_{cm}$ are the transverse mass of
the produced particles and the rapidity distributions. $m_{J/\Psi}
\simeq 3096.9\pm 0.011 MeV$ \cite{pdg} and $B_{ll'}$, the branching
ratio is for muons or electrons i.e. its for lepton pairs $J/\Psi
\rightarrow \mu^+\mu^-/e^+e^-$, is taken as $5.93\pm
0.10\times10^{-2}$ \cite{pdg} in calculating the above equations.
\par
In a similar fashion, the total inclusive cross-sections for the production of $J/\Psi$-mesons in different rapidities (i.e.$|y|<0.75$ and $2.5<|y|<0$
respectively) in $p+p$ collisions at $\sqrt{s_{NN}}$=7 TeV at
LHC would be,
\begin{equation}\displaystyle{ \frac{d^2\sigma}{dp_Tdy}|_{p+p\rightarrow
J/\Psi+X} = 78.54 p_T^{-3.135} \exp[-0.013(p_T^2+9.61)]
~~~~ for ~~|y|<0.75,}
\end{equation}
and
\begin{equation}\displaystyle{
 \frac{d^2\sigma}{dp_Tdy}|_{p+p\rightarrow
J/\Psi+X} = 37.52 p_T^{-3.135} \exp[-0.016(p_T^2+9.61)]
~~~~ for ~~2.5<|y|<0.}
\end{equation}
\par
In Fig. 1(a) and Fig. 1(b), we have drawn the solid lines depicting the SCM
model-based results with the help of above equations (6), (7), (8), (9)
against the experimental measurements \cite{adare}, \cite{alice} respectively.
\par
For the calculation of the rapidity distribution from the
equation (4) we can make use of a standard
relation as given below:
\begin{equation}\displaystyle{
\frac{dN}{dy}=\int \frac{1}{2\pi p_T}\frac{d^2N}{dp_Tdy}dp_T}
\end{equation}
For $p+p$ collisions, the calculated rapidity
distribution equation at RHIC-energy $\sqrt{s_{NN}}$ = 200 GeV is
\begin{equation}\displaystyle{
\frac{dN}{dy}|_{p+p\rightarrow
J/\Psi+X}=1.215\times10^{-6}\exp(-0.23\sinh{y_{cm}}),}
\end{equation}
In Fig. 2 we have plotted the rapidity distributions for
$J/\Psi$-production in $p+p$ collisions at $\sqrt{s_{NN}}$ = 200 GeV.
Data in the figure are taken from Ref. \cite{adare4} and the
line shows the SCM-based output.
\par
In a similar fashion, the SCM-based rapidity distribution equation at LHC-energy $\sqrt{s_{NN}}$ = 7 TeV in $p+p$ collisions has been given hereunder
 \begin{equation}\displaystyle{
\frac{dN}{dy}|_{p+p\rightarrow
J/\Psi+X}=8.025\exp(-0.043\sinh{y_{cm}}),}
\end{equation}
In Figure 3, we have drawn the solid lines depicting the
SCM-based results with the help of above equation (12) against the experimental measurements \cite{alice11}.
\subsection{Invariant Yields in $d+Au$, $Cu+Cu$ and $Au+Au$ Collisions at RHIC-energy $\sqrt{s_{NN}}$ = 200 GeV}
From the expression (4), we arrive at the invariant yields for the
$J/\Psi$-production in $d+Au\rightarrow J/\Psi+X$ reactions for mid
and forward-rapidities.
\begin{equation}\displaystyle
\frac{1}{2\pi p_T}  \frac{d^2N}{dp_Tdy}|_{d+Au\rightarrow J/\Psi+X}
= 7.25\times10^{-7} p_T^{-0.629} \exp[-0.13(p_T^2+9.61)]~~~~ for ~~|y|<0.35,
\end{equation}
and
\begin{equation}\displaystyle
\frac{1}{2\pi p_T}  \frac{d^2N}{dp_Tdy}|_{d+Au\rightarrow J/\Psi+X}
= 4.25\times10^{-7} p_T^{-0.629} \exp[-0.16(p_T^2+9.61)]~~~~ for ~~1.2<|y|<2.2.
\end{equation}
In Figure 4, we have drawn the solid lines depicting the
SCM-based results with the help of above two equations (13) and
(14) against the experimental measurements \cite{adare3}.
\par
For the case of $Cu+Cu$ most central collisions (0-20$\%$) at RHIC, the SCM-based calculated theoretical invariant yields for the rapidities $|y|<0.35$
and $1.2<|y|<2.2$ are given by the following equations respectively;
\begin{equation}\displaystyle
\frac{1}{2\pi p_T}  \frac{d^2N}{dp_Tdy}|_{Cu+Cu\rightarrow J/\Psi+X}
= 6.40\times10^{-5} p_T^{-0.829} \exp[-0.13(p_T^2+9.61)]~~~~ for ~~|y|<0.35,
\end{equation}
and
\begin{equation}\displaystyle
\frac{1}{2\pi p_T}  \frac{d^2N}{dp_Tdy}|_{Cu+Cu\rightarrow J/\Psi+X}
= 6.38\times10^{-5} p_T^{-0.829} \exp[-0.16(p_T^2+9.61)]~~~~ for ~~1.2<|y|<2.2.
\end{equation}
The experimental results for the invariant yields of $J/\Psi$
production as a function of transverse momenta for $Cu+Cu$ collisions are taken from Ref. \cite{adare2} at
centrality $0-20\%$ and are
plotted in Fig. 5.  The solid lines in the figure show the
SCM-induced results.
\par
Similarly, for $Au+Au$ collisions at $\sqrt{s_{NN}}$=200 GeV at
RHIC, the equations of transverse momenta spectra for
0-20$\%$ centrality regions are given by the undernoted relations.
\begin{equation}\displaystyle
\frac{1}{2\pi p_T}  \frac{d^2N}{dp_Tdy}|_{Au+Au\rightarrow J/\Psi+X}
= 1.32\times10^{-3} p_T^{-1.023} \exp[-0.13(p_T^2+9.61)]~~~~ for ~~|y|<0.35,
\end{equation}
and
\begin{equation}\displaystyle
\frac{1}{2\pi p_T}  \frac{d^2N}{dp_Tdy}|_{Au+Au\rightarrow J/\Psi+X}
= 0.91\times10^{-3} p_T^{-1.023} \exp[-0.16(p_T^2+9.61)]~~~~ for ~~1.2<|y|<2.2.
\end{equation}
For calculating the values
of $N_R$, in general, we have used the values of $<N_{part}>$ from
\cite{adare11}, \cite{adare3}. In the Fig. 6, the solid lines are the plots of SCM-based
invariant yields vs. $p_T$ as described by equations(17)
 and (18) at forward and mid-rapidities for $Au+Au$ collisions, while the dotted curve in the Fig. shows results of coalescence model \cite{kahana}.
The experimental
data points in the Fig.6 for the invariant yields of $J/\Psi$
production as a function of transverse momenta at centrality values
$0-20\%$ and at the rapidities $|y|<0.35$ and $1.2<|y|<2.2$
respectively are taken from  the PHENIX Collaboration \cite{adare3}.
\subsection{The Nuclear Modification Factor $R_{AB}$}
There is yet another very important observable called nuclear
modification factor (NMF), denoted here by $R_{AA}$ which for the
production of $J/\Psi$ is defined by \cite{adare2}
\begin{equation}\displaystyle{
R_{AA}=\frac{d^2N_{J/\Psi}^{AA}/dp_Tdy}{<N_{coll}(b)>d^2N_{J/\Psi}^{pp}/dp_Tdy}.}
\end{equation}
the SCM-based results  on NMFs for $Cu+Cu$ and $Au+Au$ collisions for forward rapidities
are deduced on the basis of Eqn.(6), Eqn.(15), Eqn.(17) and Eqn. (19) and they are  given by the undernoted relations
\begin{equation}\displaystyle{
R_{AA}|_{Cu+Cu\rightarrow J/\Psi+X}=0.42p_T^{0.35},}
\end{equation}
and
\begin{equation}\displaystyle{
R_{AA}|_{Au+Au\rightarrow J/\Psi+X}=0.36p_T^{0.16}.}
\end{equation}
herein the value of $<N_{coll}(b)>$  to be used is $\approx
170.5\pm 11$  \cite{alver05} for $Cu+Cu$ collisions and for $Au+Au$
collisions it is taken as $\approx 955.4\pm 93.6$  \cite{adler042}.
\par
In Fig. 7(a), we plot $R_{AA}$ vs.
$p_T$ for $0-20\%$ central region in $Cu+Cu$ and $Au+Au$ collisions.
The solid lines in the figure show the SCM-based results against the
experimentally  measured results \cite{adare3}, \cite{adare2}. The dotted lines in the fig. represent the double Color
Filtering approach \cite{kope}. And in Fig. 7(b), we plot $R_{AA}$ vs.
$N_{part}$ for $0-20\%$ central region in  $Au+Au$ collisions at $\sqrt{s_{NN}}$ = 200 GeV  and for $Pb+Pb$ collisions at $\sqrt{s_{NN}}$ = 2.76 TeV .
The solid lines in the figure show the SCM-based results against the
experimentally  measured results from Refs. \cite{adare11} and \cite{alice211} respectively. The dashed lines in the fig. represent the Gluon Saturation approach \cite{adare11}.
\section{Summary and Outlook}
Let us first concentrate on what we have achieved here: (i) The features related to $p_T$-spectra for $J/\Psi$ production in some particle-particle and nuclear collisions at various high energies have been reproduced quite successfully. (ii) The characteristics of rapidity spectra in $pp$ collisions at TeV energies have been brought out somwhat modestly satisfactorily. (iii) The features of nuclear modification factors in $Cu+Cu$, $Au+Au$ and $Pb+Pb$ reactions at RHIC and LHC energies have been worked out with the help of the applied model. Besides, some of our model-based results have also been compared with the performances on the same observables by some competing models of `standard' variety. And these comparisons with data and the results obtained by some other models reveal that SCM-based results are in better agreement with data than the other approaches grounded on the `Standard' model ilk. And this observation is no accident or coincidence. In the past such were the recurrent observations made by us valid for many other obsevables measured in the various high statistics high energy particle and nuclear experiments.
\par
Thus, summing up our past experiences and considering the weightage of the results reported here, we are forced to comment finally that
this work essentially represents a case of paradigm shift in the
domain of particle theory, as we have eschewed the conventional
views of $c \bar c$ approach to $J/\Psi$ production in the `standard'
framework. And this is just the reflection of our radical views about the particle structure and the nature of particle collisions. Obviously we obtain the fair agreement with data on some observables without inductions of (i) any QGP concept, (ii) any prognosis of suppression or enhancement of $J/\Psi$-production. The production characteristics resemble all other hadrons.

 \newpage
 \singlespacing
 
\newpage
\begin{figure}
\subfigure[]{
\begin{minipage}{.5\textwidth}
\centering
\includegraphics[width=2.5in]{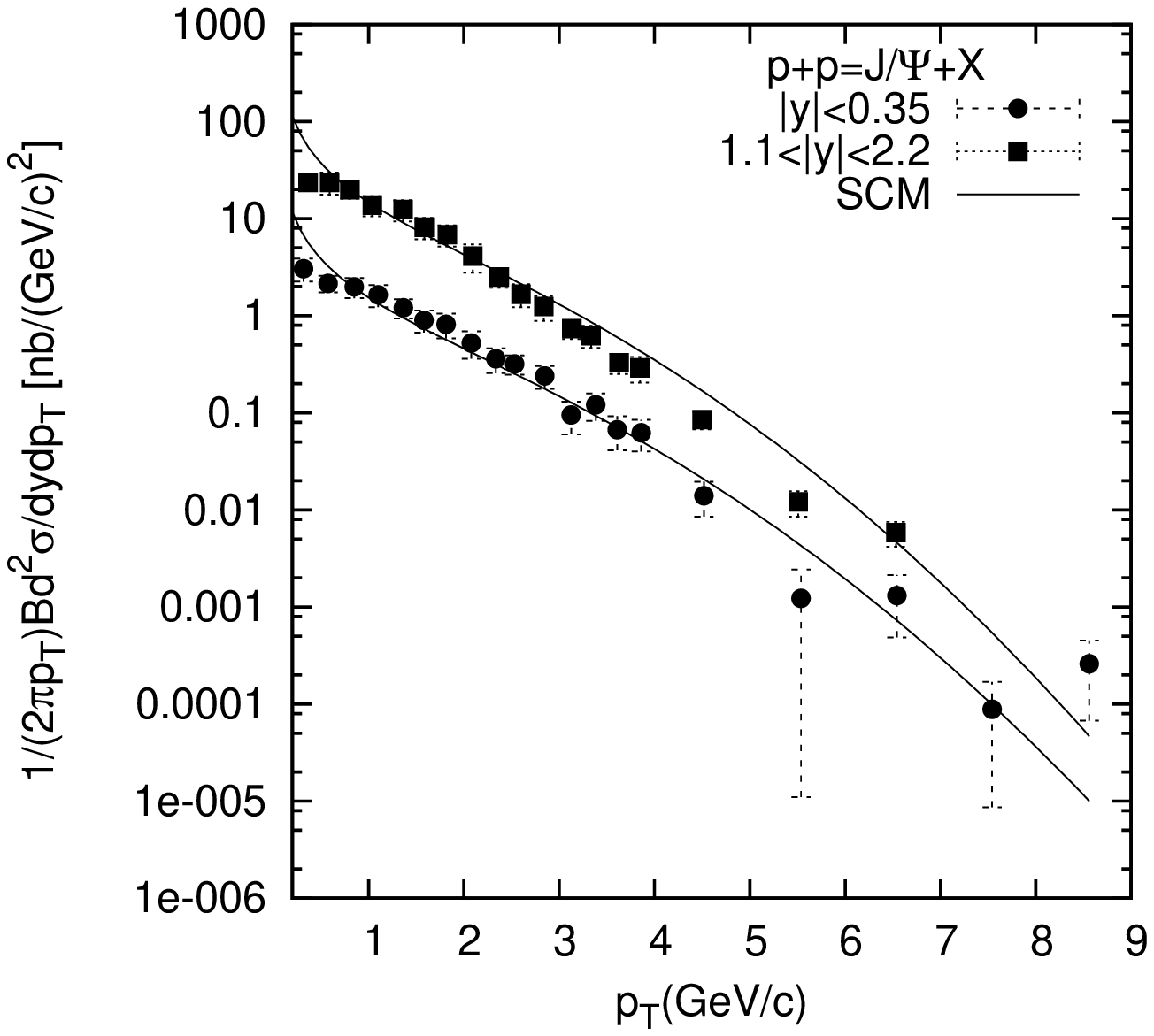}
\setcaptionwidth{2.6in}
\end{minipage}}%
\subfigure[]{
\begin{minipage}{0.5\textwidth}
\centering
 \includegraphics[width=2.5in]{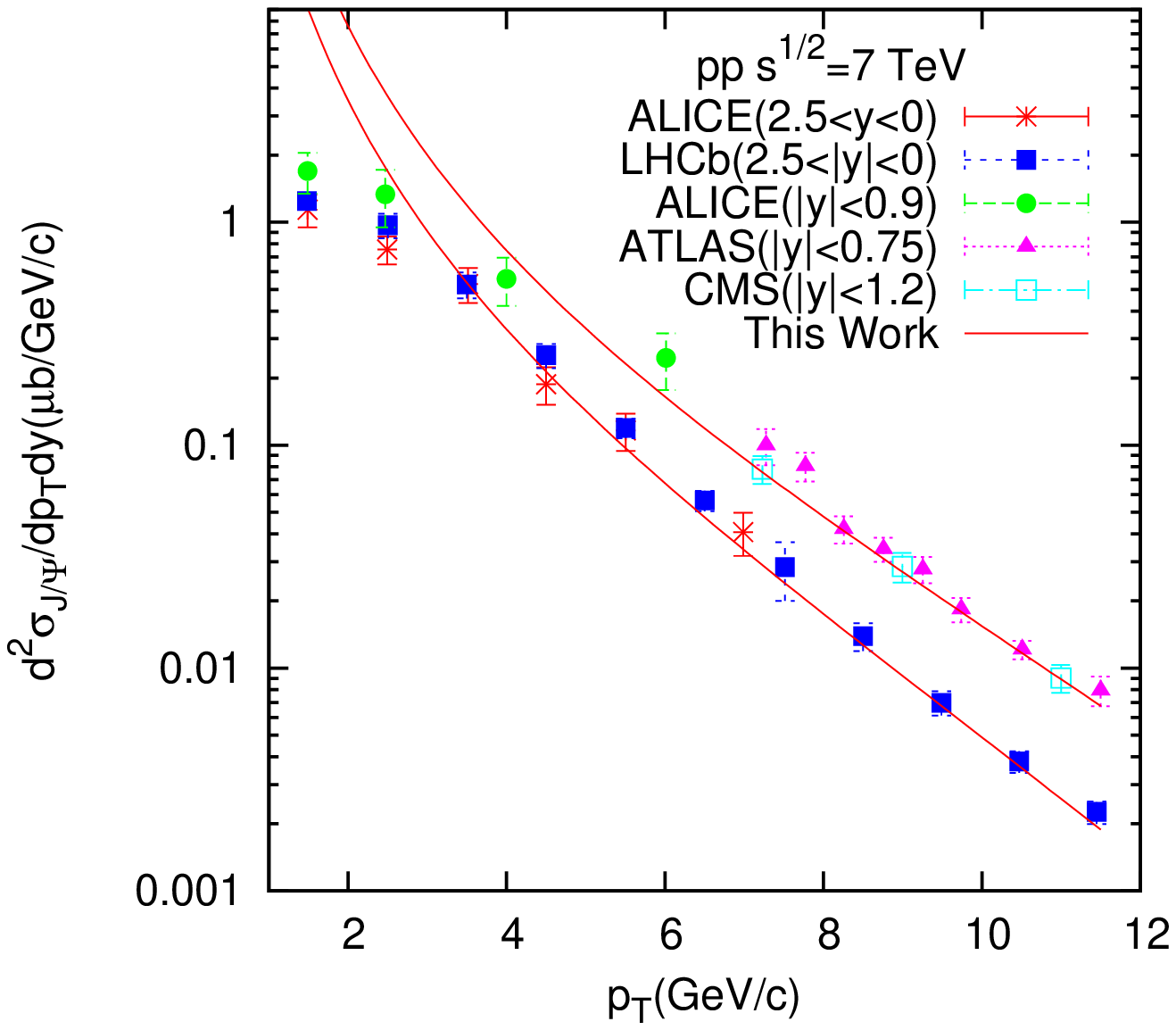}
  \end{minipage}}%
\caption{Plot of the invariant cross-section for $J/\Psi$
production in proton-proton collisions at (a) $\sqrt s_{NN} =200 GeV$ and (b) $\sqrt s_{NN} =7 TeV$ as
function of $p_T$. The data points are from \cite{adare} for (a) and from \cite{alice} for (b) . The solid
curves show the SCM-based results.}
\end{figure}
\begin{figure}
\centering
\subfigure{
\begin{minipage}{.244\textwidth}
\centering
\includegraphics[width=1.8in]{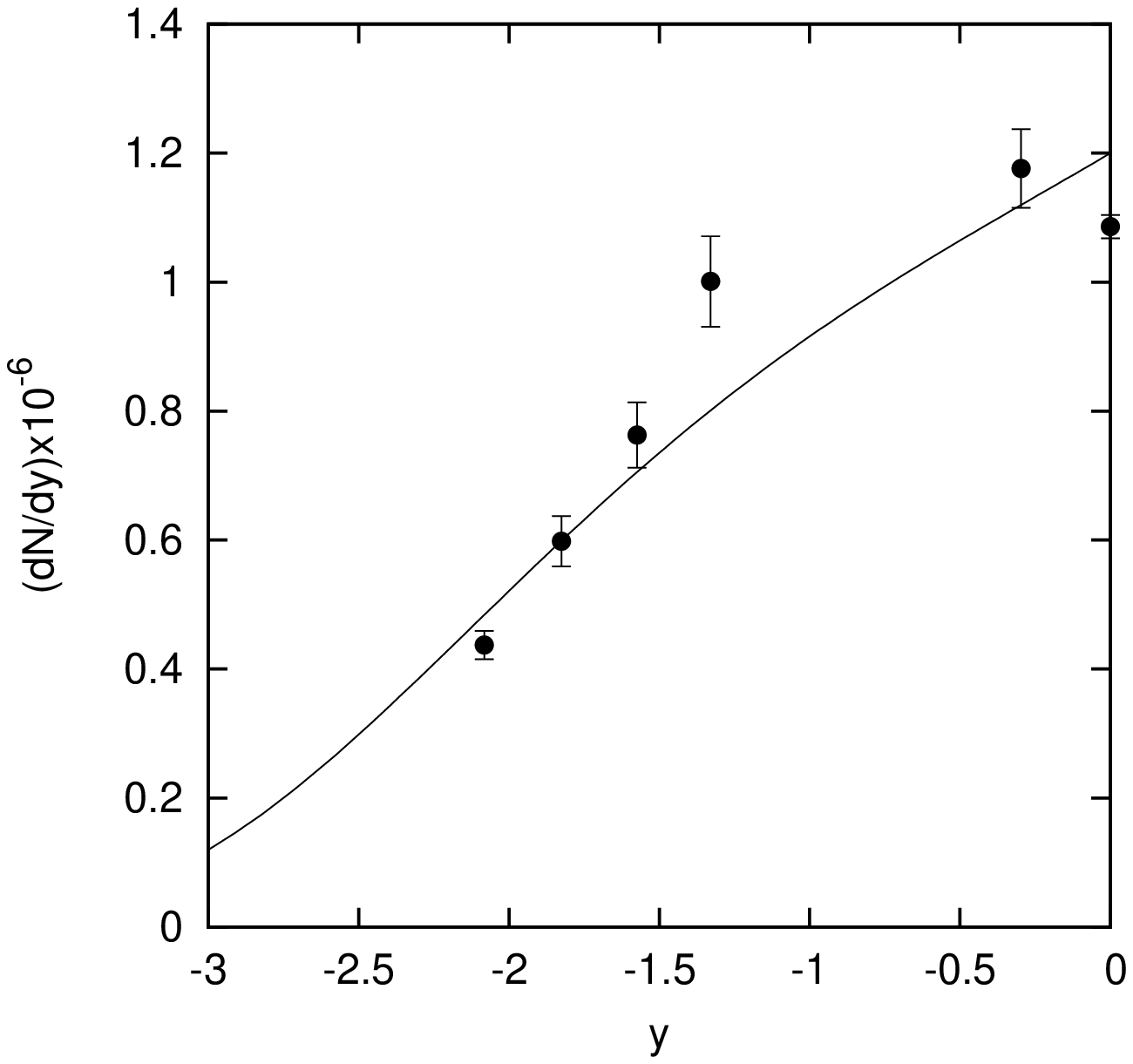}
\end{minipage}}%
\subfigure{
\begin{minipage}{0.24\textwidth}
\centering
 \includegraphics[width=1.8in]{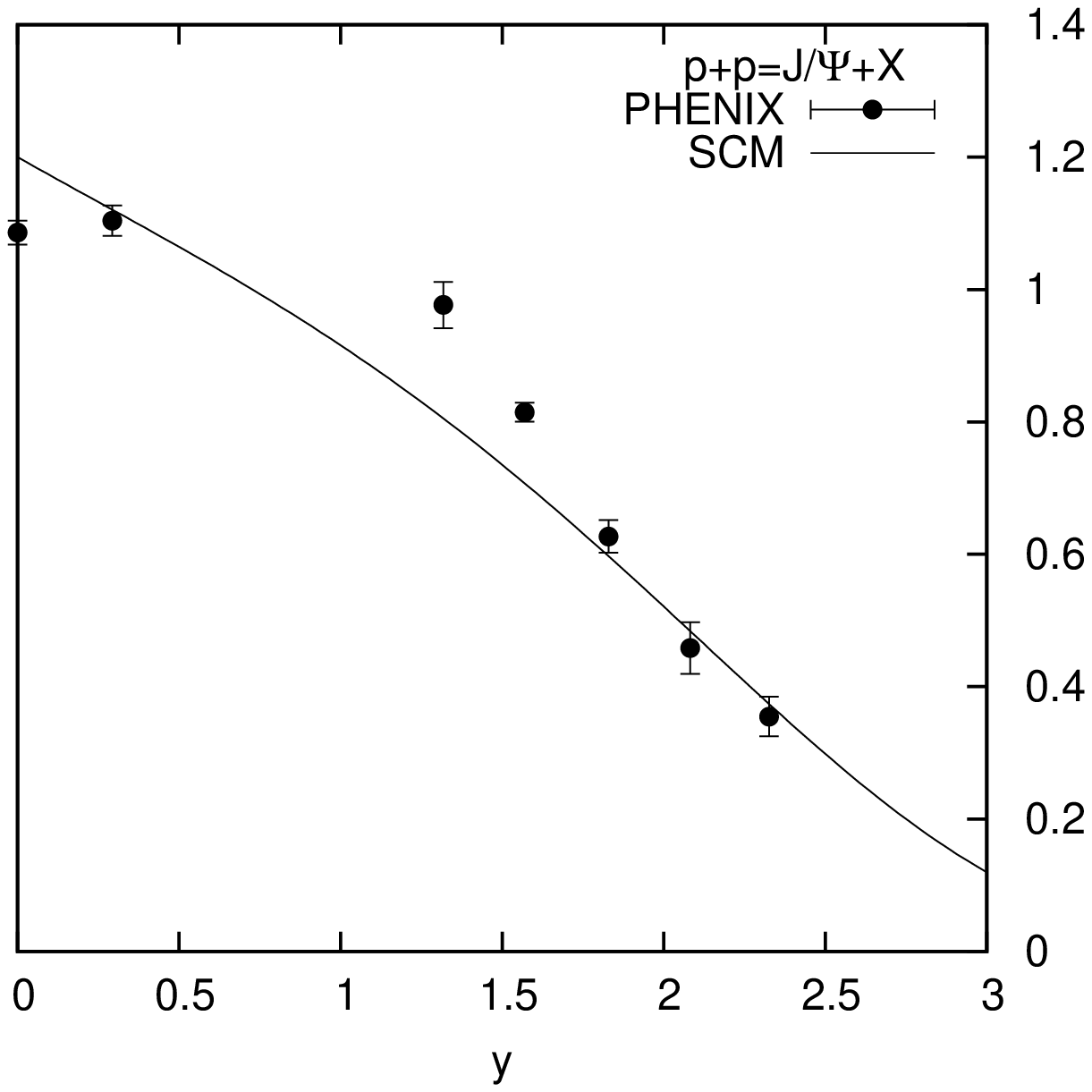}
  \end{minipage}}%
   \caption{ Plot of the rapidity distribution for $J/\Psi$ production in proton-proton
collisions at $\sqrt s_{NN} =200 GeV$ as function of $y$. The data
points are from \cite{adare3}, \cite{adare4}. The solid curves show the SCM-based
results. }
\end{figure}
\begin{figure}
\centering
\subfigure{
\begin{minipage}{.244\textwidth}
\centering
\includegraphics[width=1.8in]{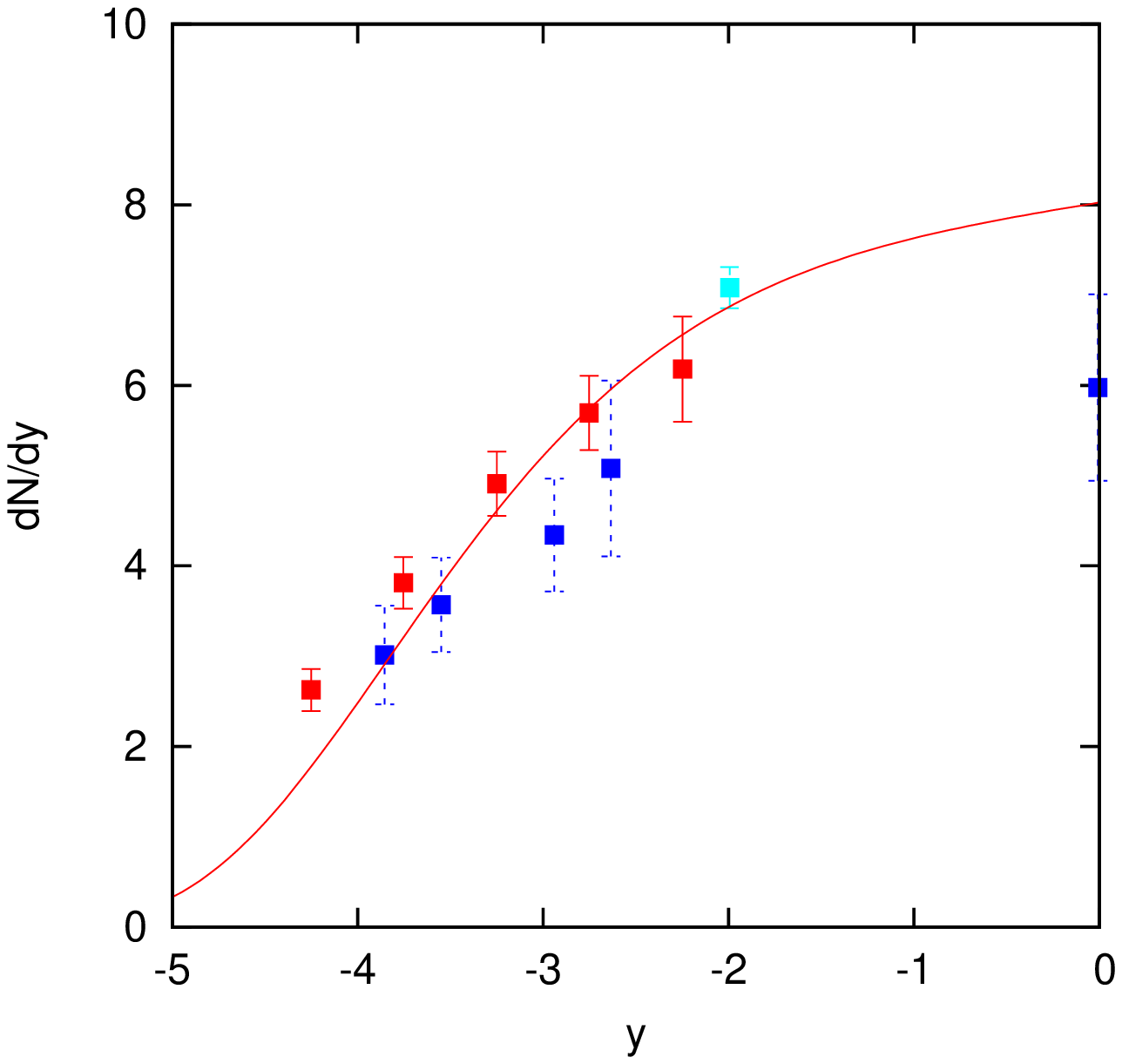}
\end{minipage}}%
\subfigure{
\begin{minipage}{0.24\textwidth}
\centering
 \includegraphics[width=1.8in]{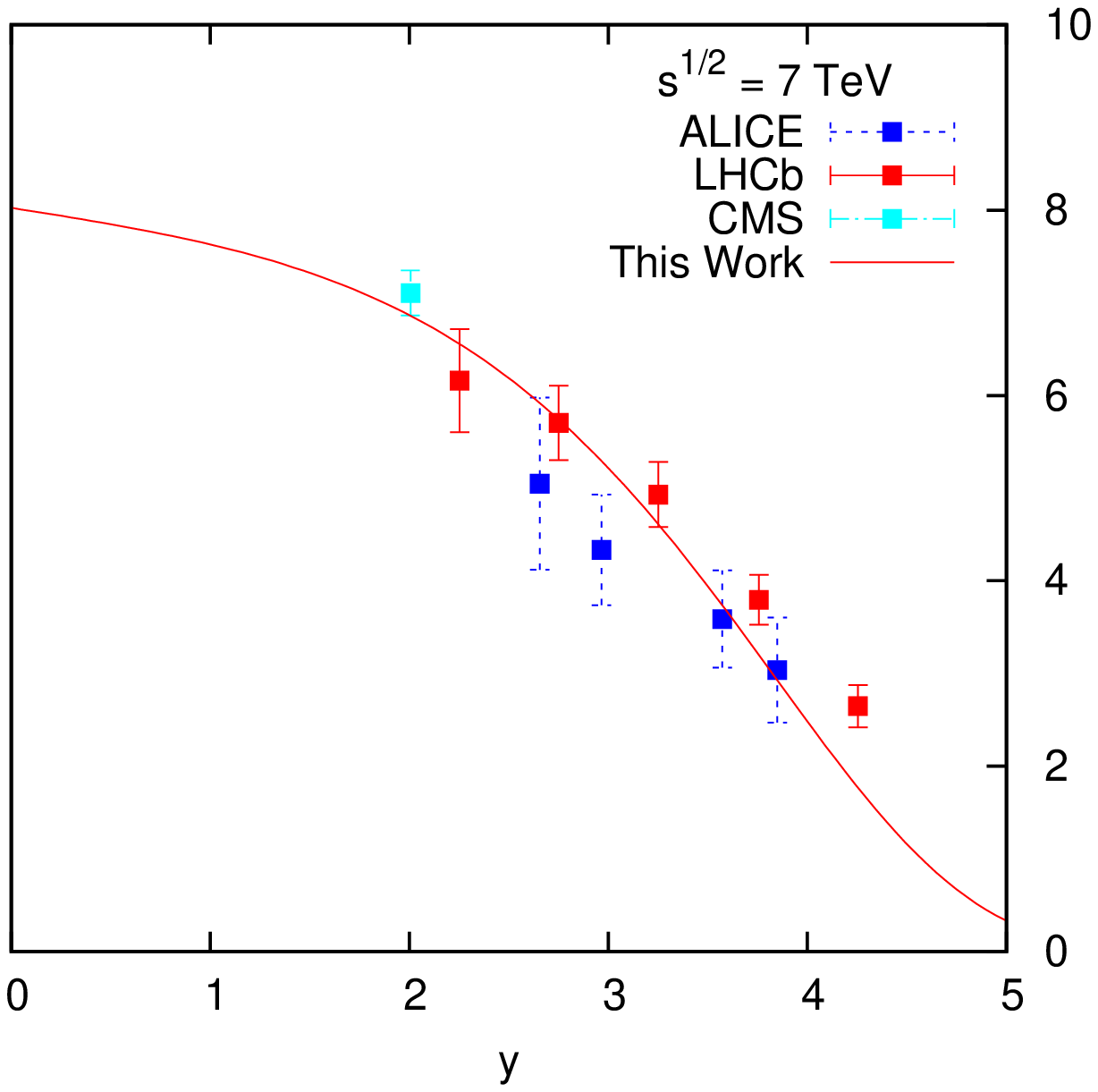}
  \end{minipage}}%
   \caption{Plot of the rapidity distribution for $J/\Psi$ production in $p+p$
collisions at $\sqrt s_{NN} =7 TeV$ as function of $y$. The data
points are from \cite{alice11}. The solid curves show the SCM-based
results.}
\end{figure}
\begin{figure}
\centering
\includegraphics[width=2.5in]{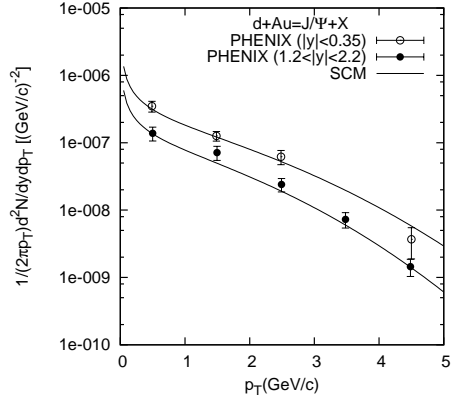}
\caption{Plot of the invariant yields for $J/\Psi$
production in $d+Au$ collisions at $\sqrt s_{NN} =200 GeV$ as
function of $p_T$. The data points are from \cite{adare3}. The solid
curves show the SCM-based results.}
\end{figure}
\begin{figure}
\centering
\includegraphics[width=2.5in]{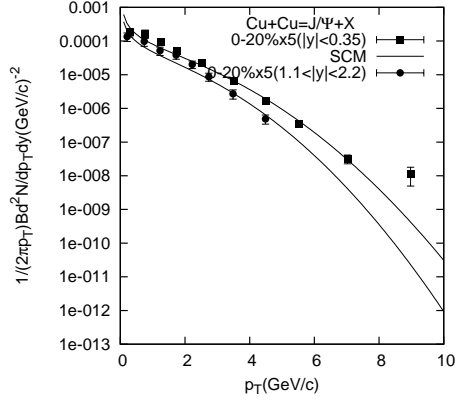}
 \caption{ Transverse momenta spectra at $|y|<0.35$ and
$1.2<|y|<2.2$ for $J/\Psi$ production in $Cu+Cu$  central collisions at
$\sqrt s_{NN} =200 GeV$. The data are taken from \cite{adare2}. The
solid curves depict the SCM-based results.}
\end{figure}
\begin{figure}
\centering
\includegraphics[width=2.5in]{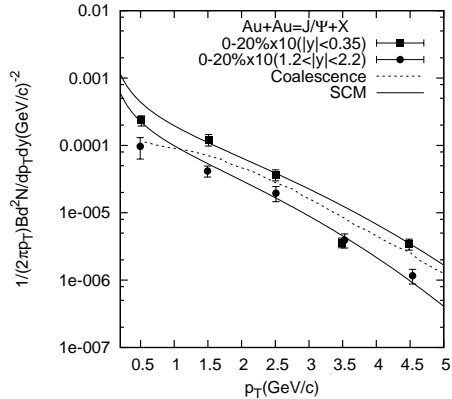}
\caption{Plot of the invariant yields for $J/\Psi$
production in $Au+Au$ collisions at $\sqrt s_{NN} =200 GeV$ as
function of $p_T$. The data points are from \cite{adare4}. The solid
curve shows the SCM-based results while the dotted curve depicts the Coalescence Model \cite{kahana}.}
\end{figure}
\begin{figure}
\subfigure[]{
\begin{minipage}{.5\textwidth}
\centering
\includegraphics[width=2.5in]{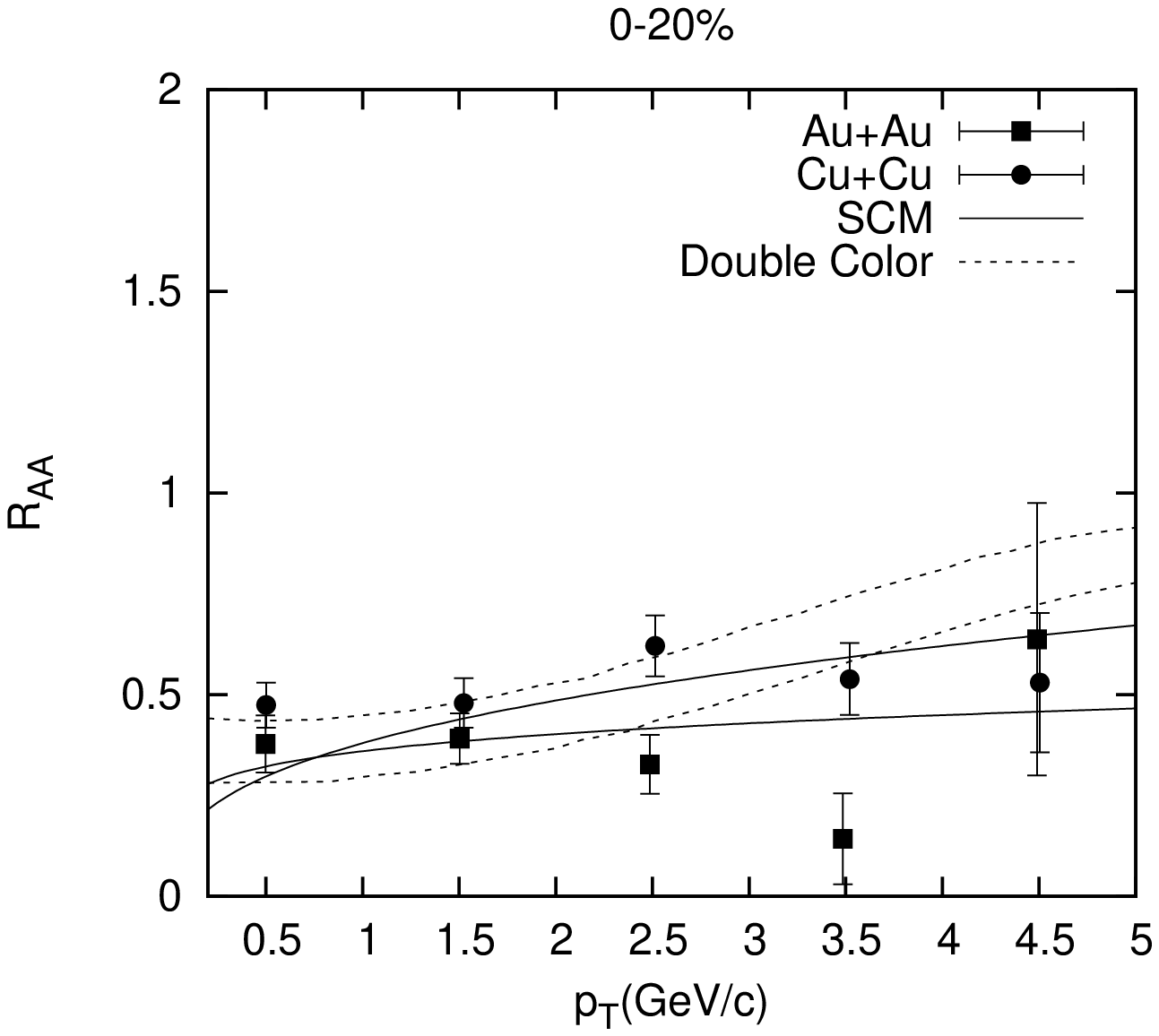}
\setcaptionwidth{2.6in}
\end{minipage}}%
\subfigure[]{
\begin{minipage}{0.5\textwidth}
\centering
 \includegraphics[width=2.5in]{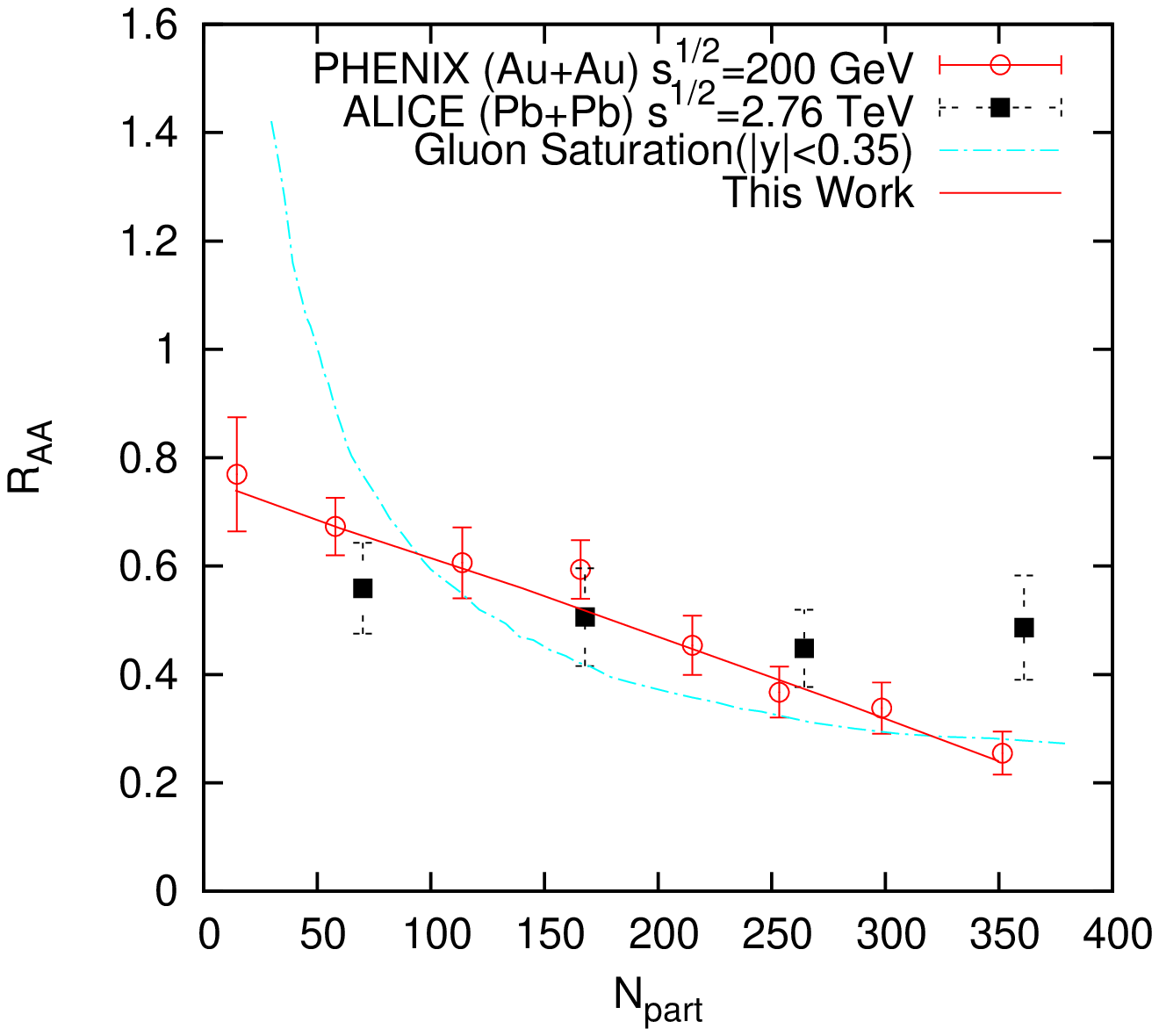}
  \end{minipage}}
  \caption{{(a)Plot of the $R_{AA}$ for $J/\Psi$ production in  $Cu+Cu$ and$Au+Au$
collisions at $\sqrt s_{NN} =200 GeV$ as function of $p_T$. The data
points are from \cite{adare3}, \cite{adare2}. The solid and dotted curves show respectively the SCM and the Double Color Filter-oriented \cite{kope}
results. (b) Plot of the $R_{AA}$ vs. $N_{part}$ for $J/\Psi$
production in $Au+Au$ collisions at $\sqrt s_{NN} =200 GeV$ and $Pb+Pb$ collisions at $\sqrt s_{NN} =2.76 TeV$. The red circles depict $Au+Au$ collisions\cite{adare11} while the black squares represent $Pb+Pb$ collisions \cite{alice211} . The solid
curves show the SCM-based results while the dotted curve depicts the Gluon Saturation Approach \cite{adare11}.} }
\end{figure}

\begin{thebibliography}{*}
 \bibitem{matsui} T. Matsui and H.Satz, Phys. Lett B{\bf{178}}, 416
(1986).
\bibitem{thews} R. L. Thews and M. L. Mangano, Phys. Rev. C{\bf{73}}, 014904 (2006).
\bibitem{pgr05} P. Guptaroy,  Bhaskar De, G. Sanyal, S. Bhattacharyya, Int. J. Mod. Phys. A{\bf{20}}, 5037 (2005).
\bibitem{pgr08} P. Guptaroy, T. K. Garain, S. Bhattacharyya, Had. J. {\bf{31}}, 451 (2008).
\bibitem{pgr09} P. Guptaroy, T. K. Garain, Goutam Sau, S. K. Biswas, S. Bhattacharyya, Had. J.
{\bf{32}}, 95 (2009). [arXiv:0906.2612 v1 [hep-ph] (15 Jun 2009)];
\bibitem{bhat881} S. Bhattacharyya, IL Nuovo Cimento C{\bf{11}}, 51 (1988).
\bibitem{pgr10} P. Guptaroy, Goutam Sau, S. K. Biswas, S. Bhattacharyya, IL Nuovo Cimento B {\bf{125}}, 1071 (2010).
[arXiv:0907.2008 v2 [hep-ph] (4 Aug 2010)].
\bibitem{pgr072} P. Guptaroy,  Goutam Sau, S. K. Biswas,  S.
 Bhattacharyya,  Mod.Phys. Lett. A.{\bf 23}, 1031 (2008).
 \bibitem{bhat882}  S. Bhattacharyya, J. Phys. G{\bf{14}}, 9 (1988).
 \bibitem{wong} C. Y. Wong:`Introduction to High-Energy Heavy Ion Collisions'
(World Scientific, Singapore, 1994).
\bibitem{adare11} A. Adare et al., PHENIX Collaboration, arXiv:1103.6269v1 [nucl-ex] (31 Mar 2011).
\bibitem{kahana} D. E. Kahana, S. H. Kahana, J. Phys. G{\bf{37}}, 115011 (2010).
\bibitem{kope} B. Z. Kopeliovich, arXiv:1007.4513v2 [hep-ph] (27 Aug 2010).
\bibitem{pdg} K. Nakamura  et al, Particle Data Group, J. Phys. G {\bf{37}}, 075021 (2010).
\bibitem{adare} A. Adare et al., PHENIX Collaboration, Phys. Rev. Lett. {\bf{98}}, 232002, (2007).
\bibitem{alice} G. M. Garc\'{i}a, arXiv:1106.5889v1 [nucl-ex] (29 Jun 2011).
\bibitem{adare4} A. Adare et al., PHENIX Collaboration, arXiv:1010.1246v1 [nucl-ex] (6 Oct 2010).
\bibitem{alice11} R. Arnaldi (for the ALICE Collaboration), J. Phys. G {\bf{38}}, 124106 (2011).
\bibitem{adare3} A. Adare et al., PHENIX Collaboration, Phys. Rev. C {\bf{77}}, 024912, (2008).
\bibitem{adare2} A. Adare et al., PHENIX Collaboration, Phys. Rev. Lett. {\bf{101}}, 122301, (2008).
\bibitem{alver05} B. Alver et al., PHOBOS Collaboration, Phys. Rev. Lett.{\bf{96}}, 212301, (2006).
\bibitem{adler042} S. S. Adler et al., PHENIX Collaboration, Phys. Rev. C{\bf{69}}, 034909, (2004).
\bibitem{alice211} P. Pillot (for the ALICE Collaboration), J. Phys. G {\bf{38}}, 124111 (2011).
\end{thebibliography}
\end{document}